\documentclass[aps,prd,showkeys,showpacs,amssymb,cite,
amsfonts,epsf,preprintnumbers,nofootinbib,superscriptaddress]{revtex4}

\usepackage[dvips]{graphicx}
\usepackage{bm,latexsym,amsmath,amssymb,amsfonts,color}
\usepackage[mathscr]{eucal}
\newcommand{\be}[1]{\begin{equation} \label{#1}}
\newcommand{\ee}{\end{equation}}
\newcommand{\bea}{\begin{eqnarray}}
\newcommand{\eea}{\end{eqnarray}}
\newcommand{\ba}{\begin{array}}
\newcommand{\ea}{\end{array}}
\newcommand{\bel}{\begin{align}}
\newcommand{\eel}{\end{align}}
\newcommand{\nn}{\nonumber}

\newcommand{\CM}{C_{V-}}
\newcommand{\A}{\mathfrak{A}}
\newcommand{\IT}{{\it inward temperature}}
\newcommand{\IHC}{{\it inward heat capacity}}

\definecolor{pptbg}{rgb}{0.961,0.945,0.863}

\begin{document}

\title{ Growth of a Black Hole on a Self-Gravitating Radiation}
\author{Hyeong-Chan Kim}
\email{hckim@ut.ac.kr}
\affiliation{School of Liberal Arts and Sciences, Korea National University of Transportation, Chungju 380-702, Korea}
%
\begin{abstract}
We feed a black hole on a self-gravitating radiation and observe what happens during the process.  
Considering a spherical shell of radiation, we show that the contribution of self-gravity makes the thermodynamic interaction through the bottom of the shell be distinguished from thermodynamic interaction through its top. 
The growth of a black hole horizon appears to be a sudden jump rather than a sequential increase.
We additionally show that much of the entropy will be absorbed into the black hole only at the last moment of the collapse. 
\end{abstract}
\maketitle


How does a black hole grow and what happens when one injects substances into it?  
What is the minimal reaction of the gravity+matter system in this situation?
Although gravity collapse has been studied extensively, gravity collapse is a nonlinear and highly dynamic process, so even in classical gravity theory, this question is far from an analytical answer.
The worse, the gravitational collapse is also believed to be closely related with the unknown quantum gravitational phenomena. 
One of the main difficulties is the uncontrollable nonlinear nature of the gravitational collapse.
To circumvent this difficulty in this work, we design a quasi-static injection process in a black hole, which consider the self-gravity of matters.  
By considering the quasi-static process, we convert the uncontrollable non-linear dynamics to a controllable one.
Evidently, a quasi-static process will minimize the entropy incremental.
In this sense, the process corresponds to the minimal reaction of the gravity.  
 
A remaining task is to build the quasi-static self-gravitating system around a black hole. 
We introduce a self-gravitating radiations bounded by two (artificial) walls located at $r= r_+$ and $r_-$. 
We, then, put part of the sphere into the black hole in succession until all the energy enter into the black hole. 
In the mean while, we keep track of the records happening on the black hole and on the matters surrounding it.
Such a self-gravitating system confined in a shell was studied recently in Ref.~\cite{Kim:2017suf}, which model is a generalization of the well-known system originally introduced by Sorkin, Wald, and Jiu~\cite{Sorkin:1981wd} in 1981 as a spherically symmetric solution which maximizes entropy.   
A self-gravitating isothermal sphere was dealt by
Schmidt and Homann~\cite{Schmidt:1999tr} where they called the geometry a `photon star'. 
The heat capacity and the stability of the solution were further analyzed in Refs.~\cite{Pavon1988,Chavanis:2007kn,Chavanis:2001hd,Chavanis:2001ib}. 
The structures of spherically-symmetric self-gravitating system were further studied in diverse area~\cite{Page:1985em,Gao:2011hh,Anastopoulos:2013xdk,Gentle:2011kv,Pesci:2006sb}.
An analytic approximation was tried in Ref.~\cite{Zurek} in analogy with the situation that a blackhole is in equilibrium with the radiations.
Recently, a complete analysis and the classification of the solutions were done in Ref.~\cite{Kim:2016jfh}. 
A star solution may have a regular or a conically-singular center. 
In Ref.~\cite{Anastopoulos:2014zqa}, the solution was argued to have an `approximate horizon', which mimics an apparent horizon~\cite{Kim:2016jfh}.

In other point of view, the model can be regarded as a self-gravitating generalization of the brick wall model, which was introduced by t'Hooft~\cite{tHooft:1996rdg} to explain the area proportionality of the black hole entropy. 
The original brick wall model is not very successful because it requires an ultra-violet cutoff.
Reasons for this failure are the ignorance of quantum gravity and disregard of the self-gravity of matter.
For the former, there is no concrete answer but various speculations. 
For the latter, correct manipulation of general relativity for the matters would do the role, which is the purpose of the present work.
First, we summarize the solution of self-gravitating radiations.
Then, we write down some of new aspects of the system especially when the radiation interacts with the inside. 
By using the results, we study the growth of a black hole on a self-gravitating radiation.

Let us explain the physical situation. 
The geometry inside the inner wall located at $r_-$ is assumed to be spherically symmetric.
To an outside observer located at $r \geq r_-$, it will be characterized by a mass parameter $M_- (< r_-/2)$ due to the Birkhoff's theorem. 
It may contain a black hole or a star, whose details are unimportant for the present purpose.
The shell between the two walls is filled with a radiation whose equation of state is $w = p/\rho= 1/3$. 
The geometry of the shell is described by the metric,
\be{ds2}
ds^2 = 
-\frac1{\beta_+^2} \sqrt{\frac{\sigma}{\rho(r)}} 
dt^2 + \frac{dr^2}{\chi^{2}} + r^2 d\Omega^2, \qquad r_-\leq r \leq r_+ ,
\ee
where $\beta_+$, $\sigma$ and $\rho(r) \equiv \sigma T(r)^4$ are the inverse temperature measured at an asymptotic infinity, the Stefan-Boltzmann constant, the local density of the radiation, respectively, and 
\be{chi}
\chi^2(r)\equiv 1- \frac{2m(r)}r;
\qquad m(r) =M_-+  4\pi \int_{r_-} ^r \rho(r') r'^2 dr'.
\ee
Then, the temperature at the outer wall is given by
$
\beta_+^{-1} = \chi_+ T_+   
$
where $\chi_+ \equiv \chi(r_+)$ and $T_+\equiv T(r_+)$.
The total mass inside the outer wall at $r_+$ is $M_+ \equiv  m(r_+)$ and the mass of the radiation in the shell is 
\be{Mrad}
M_{\rm rad}  =  M_+- M_- .
\ee
We assume that the confining walls are massless and there is no substance outside the outer wall. 
We also assume that the radiation is thermodynamically disconnected from the matter inside $r_-$ for the time being.

Introducing scale invariant variables
\begin{equation} \label{uv:r}
u\equiv \frac{2m(r)}{r}, \qquad v\equiv \frac{dm(r)}{dr} 
= 4\pi r^2 \rho(r),  
\end{equation} 
the Einstein field equation is reduced to an autonomous equation between $u$ and $v$,
\begin{equation} \label{dvdu}
\frac{dv}{du} = f(u,v) \equiv \frac{F}{G};
\qquad F\equiv 2v\Big(1-2u-\frac{2v}3\Big), \qquad G \equiv (1-u)(2v-u).
\end{equation}
Given $u_+\equiv  2M_+/r_+$ and $v_+\equiv 4\pi r_+^2 \rho(r_+)$ at the outer wall, Eq.~\eqref{dvdu} determines a unique solution curve $C_\nu$ on the $(u,v)$ plane numerically.
None of the solution curves pass the line $v=0$ and $u=1$.
Therefore, physically relevant regions are restricted to $v \geq 0$ and $u \leq 1$.
A  solution curve $C_\nu$ is characterized by the orthogonal distance of the curve from the line $u=1$,
\be{nu}
\nu \equiv  1-u_H = \chi(r_H)^2, \qquad 0\leq \nu\leq \nu_r \approx 0.50735,
\ee
where we use the subscript $H$ to denote the fact that the corresponding value is evaluated on an `approximate horizon' defined in Refs.~\cite{Anastopoulos:2014zqa,Kim:2017suf}.
A given solution curve is parameterized by a logarithmic radial coordinate
\be{xi}
\xi \equiv \log \frac{r}{r_H},
\ee
where $r_H$ is the radius of the `approximate horizon'.
Therefore, the physically relevant region on $(u,v)$ plane can be equivalently coordinated by using the set $(\nu, \xi)$. 
Eventually, an isothermal sphere of radiation of radius $r_+$ is determined by choosing a point $(u_+,v_+)$ on a curve $C_\nu$.

Spherical shell of isothermal radiations can be identified by assigning an additional point $(u_-,v_-)$ representing the inner wall at $r_-< r_+$. 
Because $(u,v)$ follows the equation of motions, the variations at the outer wall $(\delta u_+, \delta v_+)$ must be related with those at the inner wall  $(\delta u_-,\delta v_-)$.
Even though one cannot obtain the solution curve analytically in general, the variations at the inner wall are related with those at the outer wall:
\bea
\bar{\delta}u_-&=& \frac{f_+^2}{1+f_+^2}\left( \frac{B_-}{B_+}
 		+\frac{1}{f_+f_-} \right) 
		\bar{\delta} u_+
		 - \frac{1}{1+f_+^2}
	\left( \frac{B_-}{B_+}-\frac{f_+}{f_-}\right)
		\bar{\delta} v_+ 	
		 \label{delta u-} , \\
\bar{\delta}v_- 
 &=& -\frac{f_+^2}{1+f_+^2}\left( \frac{B_-}{B_+}
 			-\frac{f_-}{f_+} \right)
			\bar{\delta}u_+	
     + 	 \frac{1}{1+f_+^2}\left( \frac{B_-}{B_+}
 					+f_+f_-\right)
				\bar{\delta}v_+ 	
				, 
				 \label{delta v-} 	
\eea
where $\bar{\delta}u\equiv \chi^2\, \delta u/F$, $\bar{\delta}v \equiv \delta v/(2v-u)$, and  $B_\pm \equiv B_\pm(u_\pm, v_\pm)$ is given by 
\be{B:fA}
 B(u,v) = \sqrt{\frac{r_H}{r}} \frac{v^{3/4} \chi^3 }{
		F^2+ G^2}  .
\ee
Here $r/r_H = e^\xi$ is implicitly dependent on $u$ and $v$.
To simplify the equation, we have modified the definition of $B$ from that in Ref.~\cite{Kim:2017suf}.
The inverse equation can be obtained once we perform $+ \leftrightarrow -$.

Let us consider a heat flow $\delta M_\pm$ through the outer/inner wall.
We would like to show that the heat flow through the inner wall appears to follow different rule from that through the outer wall.
First, we would like to argue that the entropy variation presents two distinguished temperature-like quantities, each are defined for the interaction through the inner/outer wall.
As in Ref.~\cite{Kim:2017suf}, the entropy variational relation is 
\be{dS}
\delta S_{\rm rad} = \beta_+ \delta M_+ -\beta_- \delta M_- = \beta_+ \delta M_{\rm rad} +(\beta_+ - \beta_-) \delta M_- 
= (\beta_+-\beta_-) \delta M_+ + \beta_- \delta M_{\rm rad}, 
\ee
where $S_{\rm rad}$ is the the entropy of the radiations inside the shell and
\be{beta}
\beta_-^{-1}  \equiv \chi_- T_-.
\ee
Note that, $\beta_-^{-1}$ is different from $\beta_+^{-1}.$
The local temperature at the inner boundary follows from the Tolman relation, $T_- \equiv T(r_-) = \beta_+^{-1}/\sqrt{-g_{tt}(r_-)}$.
Formally, Eq.~\eqref{dS} is different from the ordinary thermodynamic first law because the entropy variation is dependent not only on $\delta M_{\rm rad}$ but also on $\delta M_-$. 
In fact, $\beta_+^{-1}$ gains the qualification as a temperature 
when $\delta M_-=0$. 
Then, Eq.~\eqref{dS} presents the first law of thermodynamics, $\delta M_{\rm rad} = \beta_+^{-1} \delta S_{\rm rad}$.
On the other hand, when heats flow only through the inner wall, i.e. $\delta M_+ =0$ and $\delta M_- \neq 0$, Eq.~\eqref{dS} reveals another form of the first law, 
\be{dM:dS}
\delta M_{\rm rad} = \beta_-^{-1} \delta S_{\rm rad}.
\ee
Apparently, $\beta_-^{-1}$ plays the role of a temperature for the interactions.
Therefore, the variational relation~\eqref{dS} admits two distinguishing legitimate temperatures (or one temperature and one temperature-like parameter). 
Later in this work, we call $\beta_-^{-1}$  \IT~to distinguish it from the ordinary temperature $\beta_+^{-1}$.

Why do we need to identify the \IT? 
First, for a given mass transfer $\delta M_{\rm rad}$ through the inner wall, the entropy change of the system may not be the same as the expectation of the outside observer because of Eq.~\eqref{dM:dS} and its ordinary analogue,
\be{dS/dS}
\frac{\delta S_{\rm rad,-}}{  \delta S_{\rm rad,+}} = \frac{\beta_-}{\beta_+}.
\ee
Second, the stability of a system is related with the concavity of the entropy. 
Given a temperature, the signature of $\delta^2S_{\rm rad}/\delta M_-^2 = -\beta_-^2 C_{V-}$ is related with the stability and the heat capacity.

Now, let us calculate the heat capacity of the system for the heat transfer through the inner wall.
Let us assume that the mass of the radiation varies by the amount $\delta M_{\rm rad}$ with respect to $\delta \beta_-^{-1}$.
The \IHC~is defined by
\be{CR-:shell}
\CM \equiv \Big(\frac{\partial M_{\rm rad}}{\partial \beta_-^{-1}}\Big)_{M_+,r_\pm} = 
-\Big(\frac{\partial M_-}{\partial \beta_-^{-1}}\Big)_{M_+,r_\pm} .
\ee
When $\delta M_+=0$ with $r_\pm $ being fixed, we get from Eqs.~\eqref{uv:r}, \eqref{delta u-} and \eqref{delta v-}, 
\be{Clocal}
\mathcal{V}_- \equiv \Big(\frac{\delta M_{\rm rad}}{\delta T_-}\Big)_{M_+,r_\pm}
=-\frac{2r_-v_-}{T_-} \Big(\frac{\delta u_-}{\delta v_-}\Big)_{r_-,u_+} 
= \frac{2 r_- v_-}{T_-} \frac{1-\A }{\A f_- +f_-^{-1}} , \qquad \mathfrak{A} \equiv \frac{f_+B_+}{f_-B_-} .
\ee
Varying $\beta_-^{-1} = \chi_- T_- = \sqrt{1- 2M_-/r_-} T_-$, we additionally get
$\delta\beta_-^{-1} =  \frac{M_-T_-}{\chi_-^2 r_-^2}  \delta r_-  - \frac{T_-}{r_- \chi_-} \delta M_- + \chi_- \delta T_- .$
Because $\delta r_+=0=\delta M_+$ and $\delta r_-=0$, there remains only one independent variation of the physical parameters at the inner boundary.
Therefore, the variations $\delta T_-$ and $\delta M_-$ must be related.
Dividing this equation by $\delta M_-$ we find that the \IHC~is related with $\mathcal{V}$ by
$\beta_- \CM ^{-1} =  
	(T_- \mathcal{V}_-)^{-1} +(r_- \chi_-^2)^{-1}  .$ 
Therefore, we find
\bea \label{cv-:gen}
\CM &=& \frac{r_-\chi_-^2}{\beta_-^{-1}} \left(1 + \frac{\chi_-^2}{2v_- } \frac{f_-^{-1}+ \A f_-}{1-\A} \right)^{-1} .
\eea
One can compare this result with the ordinary heat capacity through the outer wall with the same way in Ref.~\cite{Kim:2017suf},
\bea \label{cv:gen}
C_{V+}  &\equiv&  \Big(\frac{\partial M_{\rm rad}}{\partial \beta_+^{-1}}\Big)_{M_-,r_\pm} 
  = \frac{r_+\chi_+^2}{\beta_+^{-1}}
	\left(-1 + \frac{\chi_+^2}{2v_+} \frac{f_+ + f_+^{-1} \A}{1- \A}\right)^{-1}
.
\eea
Note that $C_{V-}$ can be obtained from $C_{V+}$ by changing the global signature after performing $\A \to \A^{-1}$ and $+ \to -$.

To model a self-gravitating radiations surrounding a black hole of radius $r_{\rm bh}=2M_- \lesssim r_-$, we consider the case that the inner/outer boundary is located around/outside the `approximate horizon' so that $1-u_- \ll 1$.
As shown in Ref.~\cite{Kim:2016jfh}, 
in this case, the solution to Eq.~\eqref{dvdu} supported by Eq.~\eqref{uv:r} is given by\footnote{We write $1-u$ to the second order in $\varepsilon$ to get an analytic form for the entropy of the system. The formula for $r_-$ in Ref.~\cite{Kim:2016jfh} contains a minor error which is corrected here.}
\bea \label{uvpm}
&\displaystyle 1-u_- = \varepsilon \frac{(2v_-/3+1)^2}{\sqrt{2v_-}}\left[1 + \frac{1+14v/3}{\sqrt{2v}} \varepsilon \right] ,\quad 
&r_- \approx \bar{r}_H\left(1+\frac{1-2v_-/3}{\sqrt{2v_-}}\varepsilon \right), \nn \\
&\displaystyle v_+ = \frac{\varepsilon^2}{2u_+^2 (1-u_+)^2}, \quad &r_+ \approx \frac{\bar{r}_H}{u_+},
\eea
respectively, where $\varepsilon = 9\nu/16$ and $\bar {r}_H = r_H(1-2\varepsilon/3)$.
The two solutions are matched at the point $S$ satisfying $du/dv = 1$, where 
$v_S \approx \varepsilon^{2/3}/2  $.
Note that $r_- \to r_H$ when $v_- \to 1/2$.
The temperature of the system is given by $\beta_+^{-1} = (8\pi \sigma)^{-1/4} \sqrt{\varepsilon/r_H}$.
The mass of the radiation is 
\be{Mrad}
M_{\rm rad} \equiv M_+-M_- = \sqrt{2v_-} \left(\frac{2v_-}{9}+1\right) \frac{\varepsilon \bar{r}_H}{2}.
\ee
We assume that $M_{\rm rad}$ is large enough so that the area inside the inner wall belongs to the black hole once all radiation is absorbed.
By using Eq.~\eqref{Mrad}, we can replace $\varepsilon$ with $M_{\rm rad}$,
\be{ep:M}
\varepsilon  \bar{r}_H= \frac{2M_{\rm rad}}{\sqrt{2v_-}( 2v_-/9+1)} .
\ee
Based on Eq.~\eqref{uvpm}, a direct calculation shows that $\A = O(\varepsilon^3) \ll 1$.
Then, we get from Eq.~\eqref{cv-:gen}, 
\be{CM:H}
\CM^H = \frac{\varepsilon r_-}{\beta_-^{-1}} \frac{ (2v_-/3+1)^2}{\sqrt{2v_-}} 
	\simeq \mu_- \frac{M_{\rm rad}}{\beta_-^{-1}};
   \qquad	\mu_-(v_-) \equiv\frac{(2v_-/3+1)^2}{v_-( 2v_-/9+1)}.
\ee
$\mu_-$ is a quantity representing a heat per unit mass and depending only on $v_-$.
Note that the \IHC~is positive definite and is independent of the physical quantities at the outer wall. 
On the other hand, the heat capacity for the interactions through the outer wall in Eq.~\eqref{cv:gen} becomes 
$$
C_{V+}^{H} = \frac{\varepsilon r_H}{\beta_+^{-1}} 
	\frac{1}{\sqrt{2v_-}(2v_--u_-)} .
$$
The heat capacity is positive definite only when the inner wall is located inside the `approximate horizon', i.e. $v_-> u_-/2$.

Later in this work, we assume $v_-> u_-/2$ so that the system is stable.
By using Eqs.~\eqref{uv:r}, \eqref{uvpm}, and $\rho(r)= \sigma T^4$, $T= (\chi \beta)^{-1}$, we find the relation between the \IT~and the ordinary temperature: 
\be{IT}
\beta_-^{-1}  = \sqrt{\frac{r_+}{r_-}} \frac{\chi_-}{\chi_+} \left(\frac{v_-}{v_+}\right)^{1/4} \beta_+^{-1}  
	\approx 
	\left(\frac23 v_-+ 1\right) \left(1 +
		\frac{4\sqrt{2v_-}}{3 }\varepsilon \right) \beta_+^{-1}.
\ee 
The \IT~is higher than the ordinary temperature.
Now, by using $S_\pm =r_\pm \beta_\pm/3 \times( 2v_\pm /3+ u_\pm) $ and Eqs.~\eqref{uvpm}, \eqref{ep:M}, and \eqref{IT}, the entropy of the radiation becomes
\be{S:rad}
S_{\rm rad} \equiv S_+-S_-
	\simeq
	\frac{4\beta_- M_{\rm rad}(2v_-/3+1) }{3(2v_-/9+1)}
	\simeq \frac{16\pi \sigma r_H\sqrt{2M_{\rm rad}}}{3} \frac{(2v_-)^{1/4}}{\sqrt{2v_-/9+1}}.
\ee

Let  some heat $\delta M (>0)$ fall through the inner wall to the black hole. 
We assume that the heat transfers slowly enough so that we can deal the system quasi-statistically. 
In other words, we ignore the off-diagonal part of the metric function.
From Eq.~\eqref{dM:dS}, the entropy of the system is decreased by the amount
$\delta S_{\rm rad, -}  = -\beta_- \delta M.$
On the other hand, if the same amount of energy is released through the outer wall, the entropy change will be $\delta S_{\rm rad,+} = -\beta_+ \delta M$.
The ratio of the two is, as in Eqs.~\eqref{dS/dS} and \eqref{IT}, $\delta S_{\rm rad, -}/\delta S_{\rm rad,+}= (2v_-/3+1)^{-1}$.
For large $v_-$, this ratio can be large.
Then, the radiation in the shell may have much higher information than 
expected by an external observer.
Now let us mimic the gravitational collapse using a successive drop of infinitesimal mass $ \delta M $ through the inner wall.
For the time being, we omit the subscript $-$ from the (intermediate) physical parameters at the inner wall i.e.  $u$, $v$, and $\beta$ represent the (intermediate) values at the inner wall.
Instead, $u_-$, $v_-$, and $\beta_-$ represent the initial values before the collapse.
A drop of $\delta M$ reduces the mass of the radiation to $M_{\rm rad} - \delta M$ and increases the mass inside the inner wall to $M+ \delta M$.
From the definition of the \IHC~\eqref{CR-:shell} and Eq.~\eqref{CM:H}, the variation of \IT~satisfies
\be{dT}
\beta\delta \beta^{-1} = \beta (\CM^{H})^{-1} \delta M_{\rm rad} 
= -\frac{1}{\varepsilon r_H} 
	\frac{\sqrt{2v}}{(2v/3+1)^2} \delta M
= - \frac{1}{\mu(v)} \frac{\delta M}{M_{\rm rad}} <0.
\ee
Now, from Eqs.~\eqref{uv:r} and \eqref{Clocal},  the variations of the scale invariant variables at the inner wall are related by 
\be{du,dv} 
\delta v = 16\pi \sigma r^2 T^3 \delta T
	= \frac{4 v}{ T \mathcal{V}} 
		\delta M_{\rm rad}
	=\frac{2v-1}{2v (2v/3+1)} \,(1-u)\delta u ,
\ee
where we use $\delta u = 2\delta M/r$ and 
$
T\mathcal{V} \approx 2r v f \approx 4rv^{2}(2v/3+1)/[(1-u) (1-2v)] .
$
Note that $u$ tends to increase but $v$ tends to recede from $v= 1/2$ during the collapse. 

We notice that $\delta v \ll \delta u$ because of $(1-u) \propto \varepsilon^2$ factor located in the right-hand-side of Eq.~\eqref{du,dv}. 
Therefore, we may say that the density at the bottom practically does not change during the drops. 
Now, let us parameterize the fallen and the remaining masses of the radiation as follows: 
\be{DM}
M_{\rm rad}(\alpha) = (1-\alpha)M_{\rm rad,0}, \qquad 
\Delta M = \alpha M_{\rm rad,0}, \qquad 0\leq \alpha \leq 1 ,
\ee
where$M_{\rm rad,0}$ is the initial mass of the radiation in Eq.~\eqref{Mrad} and $M_{\rm rad}(\alpha)$ represents the remaining mass after the $\alpha$ fraction of the initial radiation has fallen.
At the end ($\alpha =1 $), $u$ arrives at one, which implies the formation of a black hole.
Setting $v(\alpha) = v_-$, the temperature at the time of $\alpha$ fraction fallen can be found from Eq.~\eqref{dT}:
\be{beta:alpha}
\beta(\alpha)^{-1} = (1-\alpha) ^{\mu_-^{-1}} \beta_-^{-1}.
\ee
Note that the temperature of the radiation monotonically decreases to zero.
Now, let us see the entropy of the radiation during the collapse.
As in Eq.~\eqref{S:rad}, the entropy of the system becomes 
\be{S:alpha}
S_{\rm rad}(\alpha) = \frac{4\beta_+(\alpha) M_{\rm rad}(\alpha)}{3(1+ 2v/9)} 
	\approx S_{\rm rad,0} (1-\alpha)^{1-\mu_-^{-1}},
\ee
where $S_{\rm rad,0}$ represents the initial entropy before the collapse.
Note that much of the entropy is absorbed into the black hole only at the last moment because $0<1-\mu_-^{-1} < 1/2$ for  $v_->3/2$.
The inner wall will not be included in the black hole horizon until the last moment of the collapse. 
Only at the last moment when the last piece of the radiation falls through the inner wall, the whole area corresponding to $v_-> 0$ is included into the black hole, i.e the event horizon increases to encompass the inner wall. 
 So in a sense, the formation of an event horizon is a sudden leap rather than a sequential process. 

\vspace{.1cm}
In summary, we have studied the (classical) growth of a black hole by absorbing a spherical shell of self-gravitating radiations.
We have shown two important facts. 
First, the {\it inward temperature} describing thermodynamic interactions of the shell with the interior is higher than that with the outside, where the details of the interior are not important.
The {\it inward temperature} is different from the Tolman temperature and reflects the effect of self-gravity of the radiation and the curvature effect.
For a given energy transfer from the radiation, the information transmission through the inner wall is suppressed by the ratio between the ordinary to {\it inward temperature}.
Second, during the gravitational collapse, the temperature of the radiation monotonically decreases to zero.
At the time the temperature vanishes the horizon forms.
Because of this, much information will be absorbed into the black hole only at the last moment of the collapse. 

From the experience of the brick wall model, one usually expect that the density of particles at the bottom diverges when it hits the surface of a black hole. 
Such case corresponds to the $v_-\to \infty$ limit of the present case. 
For other cases, the density at the bottom takes a finite value even at the end of the collapse.
A biggest difference from the brick wall model is that the present model includes the self-gravity of the radiations rather than considering only the background gravity of a black hole. 
Notice that the density varies greatly at the thin region around $r_-\sim 2M_-$.
As given in Ref.~\cite{Kim:2016jfh}, the solution curve follows Eq.~\eqref{uvpm} for $\varepsilon^{2/3}< v_-< \varepsilon^{-2/3}$.
In this range, the radius is changed only by $O(\varepsilon^{2/3})$.
There is no curvature singularity too because the curvature components at the bottom of the box must be in order of $O(v_-/r_-^2)$.

According to the present model, at the end of the absorbing process of the radiation, the derivative of the entropy in Eq.~\eqref{S:alpha}  with respect to the mass diverges.
Because the black hole entropy changes only by a finite amount for the mass change, the divergence implies a violation of the second law of thermodynamics.  
This is an expected signal which requires the introduction of quantum physics at the end of the classic gravitational collapse.
In other words, the growth model of a black hole should take into account the equilibrium between the radiation and the black hole's Hawking radiation. 
Let us see what the second law says.
The second law of thermodynamics requires that the entropy of the radiation + the entropy of the black hole should not decrease, i.e.
$d(S_{\rm rad} + S_{\rm bh})/d\alpha \geq 0 $, which presents 
$$
 1-\alpha \geq \left[(1-\mu_-^{-1}) \frac{2 M_{\rm bh}}{M_{\rm rad,0}} \frac{S_{\rm rad,0}}{S_{\rm bh}}\right]^{1/\mu} 
 =c_-  \left[
	\frac{M_{\rm bh} \beta_-}{2S_{\rm bh}}  \right]^{1/\mu}; 
\quad c_- \equiv\left(\frac{16}{3} \frac{1+v_-/3+ 2v_-^2/9}{(1+ 2v_-/3)(1+2v_-/9)} \right)^{1/\mu} ,
$$
where $M_{\rm bh}$ represents the mass of the black hole.
Note that $c_-$ is an $O(1)$ number for all $v_-$.
The term in the square bracket is, by using $S_{\rm bh} = 4\pi M_{\rm bh}^2$,  nothing but the ratio of the Hawking temperature relative to the radiation's \IT: $T_H/\beta_-^{-1}$. 
This implies that when the \IT~of the radiation is close to the Hawking's,  the quantum effect should be taken into account.
An important difference from the existing expectation is that the \IT~is a main concern rather than the ordinary temperature.
The ordinary temperature can be lower than the Hawking's because of Eq.~\eqref{IT}.

A problem caused by considering the quantum effect is that the Hawking radiation can interfere with the final absorption process, where the temperature goes to zero. 
However, as noticed in Ref.~\cite{Anastopoulos:2014zqa}, the self-gravitating radiation cannot be continuously connected to a static black hole unless the Einstein field equation breaks down on a macroscopic near-horizon shell. 
Let us assume that the quantum nature is described by a field $\phi$. 
In the zero temperature limit, the kinetic energy term will goes to zero which implies $\dot \phi  \to 0$.
Then, the field may take the form of a static non-radiating field similar to that of the Coulomb field.
In this case, the radial pressure becomes negative of the density and a natural stable static configuration of matters with a black hole exists as shown in Ref.~\cite{Cho:2017nhx}.

\section*{Acknowledgment}
This work was supported by the National Research Foundation of Korea grants funded by the Korea government NRF-2017R1A2B4008513.
The author thanks to APCTP.


\end{document}